\documentstyle[prb,multicol,aps,amsmath,amssymb,epsf,epsfig]{revtex}
\newcommand{\ad}{\alpha_D}
\newcommand{\vf}{v_F}

\newcommand{\hbdg}{\cal{H}_{\mathrm{BdG}}}

\newcommand{\heff}{{\cal H}^{eff}}
\newcommand{\hod}{{\cal H}_{1d}}

\title{\textit{d}-wave Quasiparticles in the Tilted Vortex Lattice}
\author{Michael A. Hermele$^1$ and Luca Marinelli$^2$}
\address{$^1$Department of Physics, University of California, Santa 
Barbara, CA 93106 \\
$^2$Department of Physics, Harvard University, Cambridge, MA 02138}

\date{\today}

\begin{document}
\maketitle

\begin{abstract}
We investigate the quasiparticle spectrum of a $d$-wave superconductor in 
the mixed state, focussing on the effects of varying the in-plane angular 
alignment (or tilting) of the vortex and crystal lattices. Our approach
starts with a linearized Bogoliubov-de Gennes equation cast into a simple
form by the Franz-Te\v{s}anovi\'{c} singular gauge transformation. The zero
magnetic field spectrum is characterized by the anisotropy $\ad$ near the four
Fermi-surface nodes; at large $\ad$ the model simplifies into a one-dimensional
form that provides asymptotic results. Numerical and analytical results
are presented and compared with previous studies. We find that tilting can be
understood in terms of the one-dimensional model at large $\ad$. We see a striking
dependence on the choice of singular gauge, consistent with earlier results, but do
not attempt to resolve the issue of which gauge best describes the
microscopic physics.

\end{abstract}

\pacs{PACS numbers: 74.60.Ec, 74.72.-h}

\begin{multicols}{2}

\begin{figure}
\epsfxsize=2in
\centerline{\epsffile{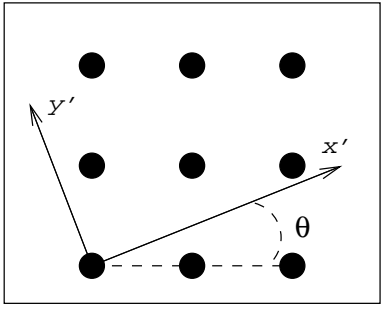}}
\caption{Schematic view of the intersection of the flux line (vortex)
lattice with a single Cu-oxide plane. The magnetic field is perpendicular to
the plane, which has an (approximately) square crystal lattice with
its orientation described by the $x'$ and $y'$ axes. The crystal axes are tilted
by an angle $\theta$ from the square lattice of vortex cores, which are represented
by the solid circles.
Previous work has generally restricted attention to
$\theta = 0^\circ$ or $45^\circ$.}
\label{tiltfig}
\end{figure}

The low-energy quasiparticle spectrum of a $d$-wave superconductor in the 
mixed state has been the focus of recent investigations. As the cuprate 
materials are believed to have an energy gap with predominately 
$d_{x^2-y^2}$ symmetry\cite{vanhar,newrmp}, such work may prove important 
for understanding certain probes of low-temperature thermodynamics and 
transport in a magnetic field. The recent studies have generally restricted 
their attention to one or two fixed vortex lattice geometries,
with or without 
small distortions to break the Bravais lattice symmetry.
However, scanning
tunneling microscopy experiments\cite{seven} on $\mathrm{YBa_2 Cu_3 O_{7 - 
\mathit{\delta}}}$ (YBCO) in a magnetic field of $6$ Tesla at $4.2$ Kelvin 
show an oblique lattice with an angle between primitive vectors of about 
$77^\circ$. This suggests that it may be important to consider
a range of 
vortex lattice geometries if a connection between experiment and the 
calculated spectra is eventually to be made. In this paper we investigate 
the low-energy spectrum as the in-plane angular alignment, or tilting, of
a square vortex lattice and the underlying crystal is varied (Fig. \ref{tiltfig}).

In zero magnetic field there are four gapless points on the Fermi surface 
of a single Cu-oxide plane, and the spectrum linearized in momentum about each 
of these points takes the form of an anisotropic Dirac cone. Using
$(x',y')$ to denote crystal lattice coordinates, throughout this 
paper we specialize to a $d_{x'y'}$ order parameter and the node at 
$(0,k_F)$, 
about which the linearized spectrum is:
\begin{equation}
E(\boldsymbol{k}) = \pm \sqrt{v_F^2 k_{y'}^2 + v_\Delta^2 k_{x'}^2} \text{.}
\end{equation}
Here $v_\Delta = \Delta_0 / p_F$, where $\Delta_0$ is the maximum 
magnitude of the superconducting gap on the Fermi surface, and
$\hbar = 1$. The anisotropy of the spectrum is characterized by 
the ratio $\ad = v_F/v_\Delta$. Experimentally $\ad$ is known to be about 
14 for YBCO and 20 for $\mathrm{Bi_2 Sr_2 CaCu_2 O_8}$\cite{quantanal}. We 
study the mixed state spectrum within the framework of the Bogoliubov-de 
Gennes (BdG) equation\cite{degennes}, generalized to incorporate a $d$-wave 
order parameter and linearized in momentum operators about the gap 
nodes\cite{simonlee,cslee,rcslee,vafek}. The Franz-Te\v{s}anovi\'{c} 
(FT) singular gauge transformation can then be used to simplify the 
problem by casting it in the form of an anisotropic two-dimensional (2d) 
massless Dirac equation with a periodic vector and scalar 
potential\cite{ft}. We will be particularly interested in the spectrum at 
large values of $\ad$, where a one-dimensional (1d) model provides 
asymptotic analytical results\cite{mel,kkb,mh} that can be compared with the 
full 2d model.

We extend the 2D model of Franz and Te\v{s}anovi\'{c}\cite{ft} to treat 
tilting at arbitrary angles. Further details of this model can be found in 
the 
studies by Marinelli, Halperin and Simon\cite{mhs} and Vafek
\textit{et. al.}\cite{vafek}; here we only 
summarize the main features.
The magnetic field satisfies $H_{c1} \ll H \ll H_{c2}$, 
implying $\xi \ll d \ll \lambda$, where $\xi$ is the Ginzburg-Landau 
(GL) coherence length, $d$ is on the order of the nearest-neighbor
vortex separation and $\lambda$ is the 
magnetic penetration depth. We remove the length scales $\lambda$ and 
$\xi$ from the problem by taking the magnetic field and the magnitude of 
the order parameter to be spatially uniform,
determining the order parameter phase by a linear GL equation. We do
not feed the resulting quasiparticle spectrum back into the BdG
equation to find a self-consistent solution, which is a good approximation 
at low temperatures. 

The heart of the FT model 
is a singular gauge transformation that requires the decomposition of the vortex lattice 
into $A$ and $B$ sublattices, so that each unit cell contains 
equal numbers of $A$ and $B$ vortices. The spectrum should be independent 
of the singular gauge used, but several studies indicate this is not the case. 
This was first pointed out by 
Vafek \textit{et. al.}\cite{vafek} and discussed further in several 
subsequent works\cite{mh,ashvin,newvafek}. We believe 
this issue may be resolved by implementing appropriate boundary 
conditions (b.c.) at the vortex cores, where the FT gauge transformation 
is not valid due to the sharp spatial variation of the order parameter magnitude.
\noindent An analysis of the physics inside the cores will likely be necessary to fix the b.c.
appropriately. 

\begin{figure}
\epsfxsize=2.5in
\centerline{\epsffile{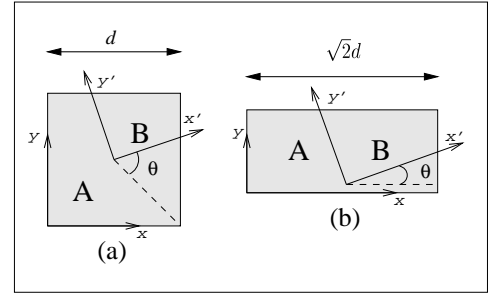}}
\caption{The magnetic unit cells for two choices of singular
gauge. The square $ABAB$ unit 
cell (a) has side $d$ and is rotated from the underlying vortex
lattice by $45^\circ$. The 
rectangular $AABB$ unit cell (b) has long side $\sqrt{2} d$ and is 
aligned with the vortex lattice. In both cases the distance between 
nearest-neighbor vortices is $d/\sqrt{2}$, and the tilting of the 
$(x',y')$ crystal coordinates is measured from the 
vortex lattice as shown. The $(x,y)$ coordinate system is aligned 
with the magnetic unit cell.}
\label{unitcells} 
\end{figure}

We do not 
attempt to address this issue here, but rather study the unregularized theory in 
two different singular gauges, each
corresponding to a different choice of sublattices. One will be called 
$ABAB$ (Fig. \ref{unitcells}a), the other $AABB$ (Fig. \ref{unitcells}b).
We work in units where $d$, the side of the $ABAB$ unit cell, is unity.
This results in the following 
linearized Hamiltonian before singular gauge transformation:
\begin{equation}
\hbdg =
\begin{pmatrix} \vf (p_{y'} - \frac{e}{c} A_{y'} ) &
\frac{1}{p_F} \{ p_{x'}, \Delta( \boldsymbol{r} ) \} \\
\frac{1}{p_F} \{ p_{x'}, \Delta^*(\boldsymbol{r}) \} &
-\vf (p_{y'} + \frac{e}{c} A_{y'} )
\end{pmatrix} \text{.}
\label{hlin}
\end{equation}
Here $\Delta(\boldsymbol{r})$ gives the 
spatial variation of the GL order parameter in the magnetic unit cell 
unit cell, which is aligned with the $x$ and $y$ axes
(Fig. \ref{unitcells}).
Upon making the FT transformation and exploiting Bloch's theorem
to write an effective Hamiltonian at every point $(k_x,k_y)$ of the first 
vortex lattice Brillouin zone, one obtains
\begin{align}
\label{htilt}
\heff(\boldsymbol{k}) = &(p_{y'} + k_{y'} + {\cal A}_{y'})\sigma_3 \\
+ \frac{1}{\alpha_D}&(p_{x'} + k_{x'} + {\cal A}_{x'})\sigma_1 + \Phi 
\text{,}\nonumber
\end{align}
where we have set $v_F = 1$.
$\boldsymbol{\mathcal{A}} = \frac{m}{2}(\boldsymbol{v}^A_s 
-\boldsymbol{v}^B_s)$ is the effective vector potential and 
$\Phi = \frac{m}{2}(v^A_{sy'} + v^B_{sy'})$ the effective scalar 
potential. 
$\boldsymbol{v}^{A,B}_s$ is the superfluid velocity associated with the 
$A, B$ sublattice.

Many of the results of Ref. \ref{mhs} still hold for this tilted
Hamiltonian. In particular, the zero energy degenerate doublet at 
the Brillouin zone center ($\Gamma$ point) is preserved to all 
orders in perturbation theory, since this result requires only 
particle-hole symmetry and a Bravais vortex lattice\cite{mhs}. Also, except 
for special tilt angles and anisotropies, we expect no zeroes at other points in the zone.
This is because tilting spoils the 
high symmetry status of the $k_y = 0$ axis where it was found there could be 
zeroes\cite{mhs}. However, invariance under 
$\boldsymbol{k} \rightarrow -\boldsymbol{k}$ is preserved.
We can therefore restrict our 
attention to the $\Gamma$ point to understand the very-low-energy band 
structure and density of states (DOS).

As pointed out by Vishwanath\cite{ashvin}, the splitting in the band 
structure near any 
degenerate doublet can be shown by first-order perturbation 
theory to be an anisotropic Dirac cone to linear order in the $k$-space 
displacement. This actually gives the exact result for the slope of the 
lowest-energy bands as the $\Gamma$ point is approached from a 
particular direction.
Writing $\boldsymbol{k} = k(\cos\phi,\sin\phi)$, we have
\begin{equation}
\label{ptham}
\heff(\boldsymbol{k}) = \heff(\boldsymbol{0}) +
k[\sin(\phi-\theta)\sigma_3 + \frac{1}{\ad}\cos(\phi-\theta)\sigma_3] 
\text{.}
\end{equation}
An elementary degenerate perturbation theory calculation gives the perturbed energies near 
the $\Gamma$ point:
\end{multicols}
\begin{equation}
\label{bss}
\epsilon_\pm =  
\pm k\sqrt{t_1 \sin^2 (\phi-\theta) + t_2 \cos^2(\phi-\theta)
+ t_3 \sin(2(\phi-\theta))} + {\cal O}(k^2) \text{,}
\end{equation}
\begin{multicols}{2}
\noindent where the coefficients can be expressed in terms of matrix elements of $\sigma_1$ and 
$\sigma_3$ between the states of the degenerate doublet.
Except for the 
orientation of the new anisotropy axes, the anisotropic Dirac cone described by
Eq. \ref{bss} is completely characterized by 
the maximum and minimum values of $\epsilon_+/k$, which we denote by $v_+$ 
and $v_-$, respectively. We will study these velocities as a 
function of $\ad$ and $\theta$.

In order to apply the 1d model to obtain information at large $\ad$,
we must specialize to rational tilt angles
$\theta = \arctan (p/q)$ for integer $p$ and $q$. 
Details of this model are 
presented in the papers of Knapp, Kallin and Berlinsky\cite{kkb}, and 
Marinelli and Halperin\cite{mh}. The essential 
idea is that as $\ad \rightarrow \infty$ the quasiparticles 
see vortices that are effectively smeared out 
in the ``hard'' $v_F$-direction, so that all gauge-invariant quantities 
become functions only of the coordinate in the $v_\Delta$-direction. 
Choosing coordinates $x$ and $y$ for the soft and hard directions, 
respectively, the 1d Hamiltonian for crystal momentum $\boldsymbol{k}$ is:
\begin{equation}
\hod(\boldsymbol{k}) =
\begin{pmatrix}
k_y + U^A(x) &  \frac{1}{\ad} (p_x + k_x) \\
\frac{1}{\ad} (p_x + k_x) &  -k_y + U^B(x)
\end{pmatrix} \text{.}
\end{equation}
Here $U^{A,B}$ is the potential encapsulating the residual effects of
the $A,B$ vortices. For the gauges considered we can always choose a 
unit cell with 
only two columns of smeared-out vortices, provided $\theta$ is rational. 
This is true because the perpendicular distance from one column of $A$/$B$ 
vortices to the nearest $A$/$B$ column is always the same.

\begin{figure}
\epsfxsize=2.5in
\centerline{\epsffile{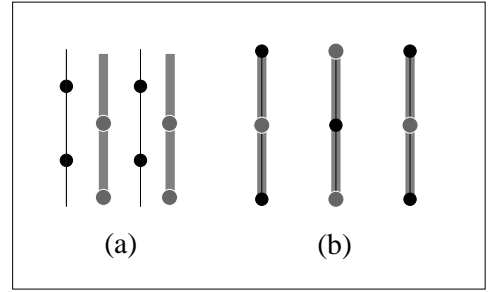}}
\caption{Diagram of the two qualitatively different configurations of the 1d model.
Dark circles/lines represent one type of vortex/column of vortices. The shaded
circles and lines represent the other. In (a) the $A$ and $B$ vortices form distinct
columns, while in (b) all columns contain equal numbers of $A$ and $B$
vortices.}
\label{1dfig}
\end{figure}

There are essentially two distinct possibilities when we go to the 1d 
model, which depend on the gauge choice and on $\theta$. These are 
illustrated in Figure \ref{1dfig}.
Configurations with distinct columns of $A$ and $B$ 
vortices (Fig. \ref{1dfig}a) were shown\cite{mh} to give rise to band structure features 
exponentially small in $\ad$. In particular, for the $ABAB$ gauge with $\theta = 
45^\circ$, it was found that $v_- \sim \exp(-\ad \pi^2 / 16)$. This can 
easily be generalized when the distance between two 
adjacent columns (of different types) is always the same, by 
scaling the 1d unit cell by a factor of $1/d_v$ in the $x$-direction and
$d_v$ in the $y$-direction, where $d_v$ is the distance between adjacent 
vortices in the same column. Following the calculation of Ref. \ref{mh} one finds
$v_- \sim \exp(-\ad \pi^2/16 d_v^2)$.

When the $A$ and $B$ columns overlap (Fig. \ref{1dfig}b)
the asymptotic behavior is
very different. It has been argued and verified numerically 
that there are in this case \textit{no} exponentially small features in 
the band structure\cite{mh}. We can go further to calculate the exact 1D 
band structure on the $k_y = 0$ line, and hence also $v_-$. Since 
$U^A = U^B = U(x)$ in this case, we have
\begin{equation}
\hod(k_x,0) = U(x) + \frac{1}{\ad}(p_x + k_x)\sigma_1 \text{.}
\end{equation}
The matrix structure of this Hamiltonian is trivial, and
it separates into two uncoupled components. The resulting differential equations are:
\begin{equation}
U(x)\psi \pm \frac{1}{\ad}(-i\frac{d}{dx} + k_x)\psi = E_\pm \psi \text{.}
\end{equation}
Upon integration one finds 
\begin{equation}
\psi (x) = \psi_0 e^{\pm i\ad \int^x (E_\pm - U(x) \mp k_x/\ad)}.
\end{equation} 
$\psi$ is a Bloch function, so we must impose periodic boundary conditions, 
thus obtaining a quantization condition 
\noindent for the energies:
\begin{equation}
\int_{-1/2d_v}^{1/2d_v}dx(E_\pm - U(x) \mp \frac{k_x}{\ad}) = \frac{2\pi 
n}{\ad} \text{.}
\end{equation}

For those solutions with $|E|$ greater than the maximum of $U(x)$, this is 
exactly the Bohr-Sommerfeld quantization condition used in Ref. \ref{mh} to obtain 
the spectrum in a WKB approximation\cite{mh}, only here it is 
exact. At the $\Gamma$ point ($k_x = 0$), there are doubly degenerate 
solutions $E_\pm = E^\Gamma_n$. The degeneracy is split for finite $k_x$ 
and $E_\pm = E^\Gamma_n \pm k_x/\ad$ (see Figs. 10 and 11 in Ref. \ref{mh}).
The lowest bands are simply $E_\pm = 
\pm k_x/\ad$, implying $v_- = 1/\ad$ in striking contrast to the case of 
separated columns. Still more striking is the fact that we can switch 
between exponential (separated columns) and algebraic 
(overlapping columns) decrease of $v_-$ simply by choosing a 
different singular gauge. For example, if the gauge choice did not 
matter the $AABB$ lattice with $\theta = 45^\circ$ would be the same 
as $ABAB$ and $\theta = 0^\circ$. But the spectra are \textit{not}
the same, since 
the former case has overlapping columns in the 1D model and the latter 
does not.  These conclusions are in excellent agreement with the numerical
results\cite{mh}.

We have numerically calculated $v_\pm$ as a function of $\ad$ and $\theta$ 
in the 2d model with a reciprocal space cutoff. Working in momentum 
space allows tilting to be handled very simply, and avoids 
the complications associated with 
discretizing a Dirac equation on a real space
lattice\cite{mhs,kogut}. The major
disadvantage is that the matrices to be diagonalized are not sparse. For 
the square unit cell of the $ABAB$ gauge we typically used a cutoff of 
$\Lambda = 10$, where $|Q_x|,|Q_y| \leq \Lambda$ and $Q_i$ are the 
coefficients of the reciprocal lattice primitive vectors. For the $AABB$ 
gauge the unit cell is non-square, so we took $2|Q_x|,|Q_y| \leq \Lambda$, 
typically with $\Lambda = 8$. It was not feasible to push $\Lambda$ much 
greater than $20$. Also, for any fixed $\Lambda$, we expect the 
numerical calculations to become less reliable as $\ad$ is 
increased, since raising the anisotropy introduces more and more 
low-energy states that do not fall within the cutoff and are improperly 
thrown out. Although the accessible range of $\Lambda$ and $\ad$ 
is not sufficient 
to provide detailed information on the true asymptotic behavior 
as $\ad \to \infty$ or
the convergence of the numerics, we believe it should be 
sufficient to extract certain important qualitative features.
\begin{figure}
\includegraphics[bb=75 35 600 705,angle=270,width=3.2in]{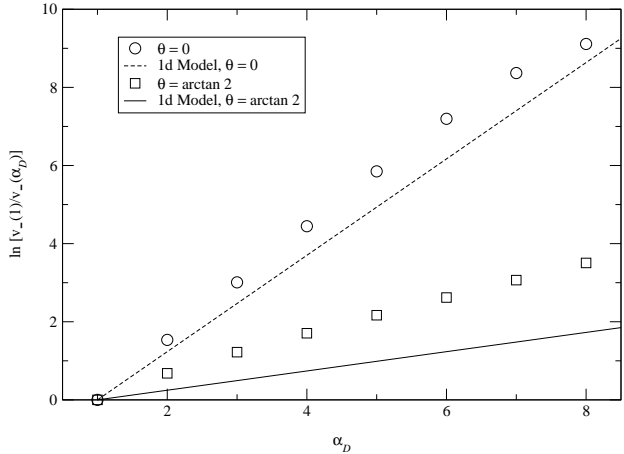}
\caption{Plot of $\operatorname{ln}(v_-(1)/v_-(\ad))$ versus the 
anisotropy $\ad$, comparing data from the 2d numerics with the 1d model. The results 
shown are in the $AABB$ gauge, with tilt angles giving distinct columns in the 1d 
model. The hollow shapes are numerical results for $\theta = 0$ and 
$\theta = \operatorname{arctan}2$. The lines are the asymptotic results 
from the 1d model and have slope $\pi^2/(16 d_v^2)$.}
\label{pureplot}
\end{figure}

\begin{figure}
\includegraphics[bb=65 35 600 705, angle=270,width=3.2in]{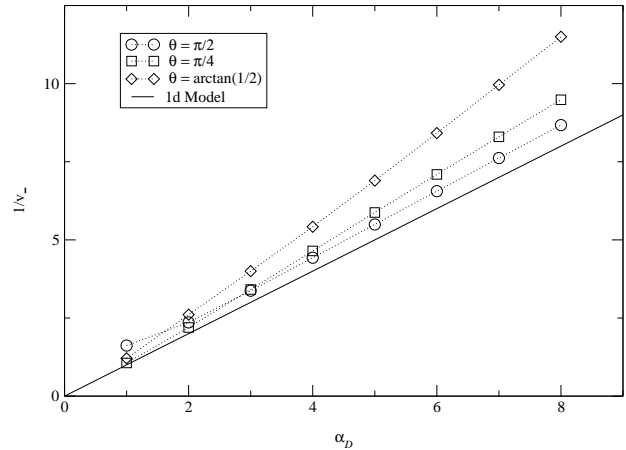}
\caption{Plot of $1/v_-$ versus $\ad$ comparing 2d numerical data in the $AABB$ 
gauge with the 1d model. The tilt angles are chosen to give overlapping columns of 
vortices in the 1d model. The shapes connected by dotted lines are 2d numerical
data for different tilt angles, and the solid
line is simply a plot of $1/v_- = \ad$, the exact 1d result.}
\label{mixplot}
\end{figure}

Our results for $v_-(\ad)$ at fixed $\theta$ show good qualitative 
agreement with the predictions of the 1d model. For tilt angles 
and gauges with distinct $A$ and $B$ columns we see roughly
exponential decrease of $v_-$ similar to that calculated analytically (Fig. 
\ref{pureplot}). We 
believe we do not see quantitative agreement because we cannot access the 
asymptotic regime numerically. When there are overlapping 
columns we see similar agreement with the analytical results;
$v_-$ is roughly proportional to $1/\ad$ with 
the constant of proportionality closer to unity when the columns are 
farther apart (Fig. \ref{mixplot}). This is 
consistent with our understanding of the 1d model; at greater column 
separation $d_v$ is smaller, and the quasiparticle wavefunctions do not 
have to be smeared out as much for them 
to be completely delocalized in the $y$-direction. Numerical calculations 
of $v_+$ show only small variations, typically by at most a factor of order unity as 
$\ad$ ranges from $1$ to $8$. Therefore the renormalized anisotropy 
$v_+/v_-$ behaves similarly to $1/v_-$ as a function of $\ad$.

We have seen that, for large $\ad$, tilting of the vortex and crystal 
lattices in the Franz-Te\v{s}anovi\'{c} framework can be understood in terms 
of 
a one-dimensional model. For tilt angles and singular gauges with distinct
columns of $A$ and $B$ vortices, $v_-$ is exponentially 
small in $\ad$, while when the columns overlap it decreases 
algebraically. In both these cases $v_+$ does not vary significantly with 
$\ad$. These results depend dramatically on the choice of singular gauge, 
and provide a further indication that the FT 
transformation must somehow be
regularized if it is to provide a consistent account of the physics. While 
proposals for the kind of regularization necessary have been discussed in the
literature\cite{mh,ashvin,newvafek}, a detailed resolution of these issues 
remains an important problem for future work.

The authors would like to thank B.I. Halperin for very useful
discussions and a careful reading of the manuscript. 
MH was supported by the Harvard MRSEC NSF/DMR-9809363 through the
REU program, by the Department of Defense with an NDSEG fellowship, and by
Regents Special and Broida Fellowships from the University of California, Santa Barbara.
LM was supported in part by the NSF grant DMR-99-81283.

\end{multicols}

\begin{references}

\bibitem{vanhar} D. J. Van Harlingen, Rev. Mod. Phys.
\textbf{67}, 515 (1995).
\bibitem{newrmp} C. C. Tsuei and J. R. Kirtley, Rev. Mod. Phys.
\textbf{72}, 969 (2000).
\bibitem{seven} I. Maggio-Aprile, Ch. Renner, A. Erb, E. Walker and
\O. Fischer, Phys. Rev. Lett. \textbf{75}, 2754 (1995).
\bibitem{quantanal} May Chiao \textit{et. al.}, Phys. Rev. B
\textbf{62}, 3554 (2000).
\bibitem{degennes} P. G. de Gennes, \textit{Superconductivity of Metals and 
Alloys} (Addison-Wesley, Reading, MA, 1989).
\bibitem{simonlee} S. H. Simon and P. A. Lee, Phys. Rev. Lett. 
\textbf{78}, 1548 (1997).
\bibitem{cslee} G. E. Volovik and N. B. Kopnin,
Phys. Rev. Lett. \textbf{78}, 5028 (1997).
\bibitem{rcslee} S. H. Simon and P. A. Lee,
Phys. Rev. Lett. \textbf{78}, 5029 (1997)
\bibitem{vafek} O. Vafek, A. Melikyan, M. Franz and Z. Te\v{s}anovi\'{c},
Phys. Rev. B \textbf{63}, 134509 (2001).
\bibitem{ft} M. Franz and Z. Te\v{s}anovi\'{c}, Phys. Rev. Lett.
\textbf{84}, 554 (2000).
\bibitem{mel} A. S. Mel'nikov,
J. Phys.: Condensed Matter \textbf{11} 4219 (1999).
\bibitem{kkb} D. Knapp, C. Kallin and A. J. Berlinsky,
Phys. Rev. B \textbf{64}, 014502 (2001).
\bibitem{mh} Luca Marinelli and B. I. Halperin, Phys. Rev. B \textbf{65},
014516 (2002). \label{mh}
\bibitem{mhs} Luca Marinelli, B. I. Halperin and S. H. Simon,
Phys. Rev. B \textbf{62}, 3488 (2000). \label{mhs}
\bibitem{ashvin} Ashvin Vishwanath,
Phys. Rev. Lett. \textbf{87}, 217004 (2001).
\bibitem{newvafek} O. Vafek, A Melikyan and Z. Te\v{s}anovi\'{c},
Phys. Rev. B \textbf{64}, 224508 (2001).
\bibitem{kogut} J. B. Kogut, Rev. Mod. Phys. \textbf{55}, 775 
(1983).
\end{references}
\end{document}